\documentclass[twocolumn,showpacs,preprintnumbers,amsmath,amssymb]{revtex4}
\usepackage{graphicx}% Include figure files
\usepackage{dcolumn}% Align table columns on decimal point
\usepackage{bm}% bold math

\begin{document}

\preprint{J. Y. Jo \textit{et al.}}

\title{Nonlinear Dynamics of Domain Wall Propagation in Epitaxial Ferroelectric Thin Films}

\author{J. Y. Jo,$^1$ S. M. Yang,$^1$ T. H. Kim,$^1$ H. N. Lee,$^2$ J.-G. Yoon,$^3$ S. Park,$^4$ Y. Jo,$^4$ M. H. Jung,$^4$}
\author{ T. W. Noh$^{1,}$}
\email{twnoh@snu.ac.kr}
\affiliation{$^1$ReCOE $\&$ FPRD, Department of Physics and Astronomy, Seoul Nat'l University, Seoul 151-747, Korea.\\
 $^2$Materials Science and Technology Division, Oak Ridge National Laboratory, Tennessee 37831,
 USA.\\$^3$Department of Physics, University of Suwon, Gyeonggi-do 445-743, Korea.\\$^4$Quantum Materials Research Team, Korea Basic Science
Institute, Daejeon 305-333, Korea.}

\begin{abstract}
We investigated the ferroelectric domain wall propagation in
epitaxial Pb(Zr,Ti)O$_3$ thin films over a wide temperature range
(3 - 300 K). We measured the domain wall velocity under various
electric fields and found that the velocity data is strongly
nonlinear with electric fields, especially at low temperature. We
found that, as one of surface growth problems, our domain wall
velocity data from ferroelectric epitaxial film could be
classified into the creep, depinning, and flow regimes due to
competition between disorder and elasticity. The measured values
of velocity and dynamical exponents indicate that the
ferroelectric domain walls in the epitaxial films are fractal and
pinned by a disorder-induced local field.
\end{abstract}

\pacs{05.45.-a,47.15.G-,64.60.F-,68.35.Rh,77.80.Dj}% PACS, the Physics and Astronomy
\maketitle

The physics of surface growth in disordered media with quenched
defects is of crucial importance to understand numerous intriguing
 natural phenomena \cite{Sornette}, including contact lines in wetting,
 surface of epitaxially grown films, and magnetic domain walls.
In such media, elastic forces tend to keep surfaces flat, while
defects locally promote the wandering, as schematically displayed
by Fig. 1(a). The competition between elastic and pinning forces
leads to a complicated energy landscape with many local minima,
which affects the surface growth dynamics under an external
force. Recently, there have been extensive reports to adapt the fractal concepts to surface growth dynamics \cite{Barabasi}.\\
%%%%%%%%%%%%%%%%%%%%[Figure 1] %%%%%%%%%%%%%%%%%%%%%%%
\begin{figure}[tbp]
\includegraphics[width=3.5in]{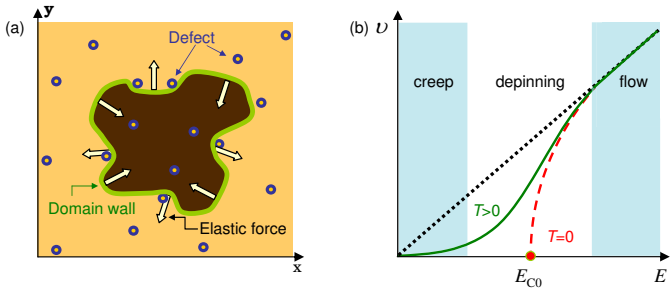}
\caption{(color online) (a) A schematic diagram of domain wall propagation in disordered medium. Elastic forces come from the curvature of domain wall, and defects work as strong pinning sites. (b) Theoretical prediction on the domain wall velocity $v$ vs. electric field $E$
in system governed by competition between disorder and elasticity effects.
$E_{C0}$ represents a threshold $E$.}
\end{figure}
%%%%%%%%%%%%%%%%%%%%%%%%%%%%%%%%%%%%%%%%%%%%%%%%%%%%%%
\indent Ferroelectric (FE) domains have been studied for past decades because of scientific importance in microscopic aspects such as
multi-domain formation, stability, and pattern at equilibrium as well as technological applicability in multi-functional devices such as FE
random access memories, actuators, and sensors \cite{Dawber, Tagantsev}. Quite recently, lots of piezoresponse force microscope (PFM) studies
have provided us microscopic aspects of FE domains, including inhomogeneous nucleation process \cite{Kim} and the fractal nature of their rough
surfaces \cite{Rodriguez}. Note that most works on FE domains have been focused on their static properties. In spite of its scientific and
technological importance, we have limited understandings on how the FE domain
wall propagates in terms of time.\\
\indent We suggest to prospect FE domain wall from the view of nonlinear
responses, which follow the predictions of the statistical physics
on surface growth. Then, the FE domain wall velocity $v$ should
have a nonlinear behavior, shown in Fig. 1(b), under $E$.  At zero
temperature $T$, the domain wall remains strongly pinned by local
disorders until $E$ reaches a threshold value $E_{C0}$. When $E
\geqq E_{C0}$, it experiences a pinning-depinning transition and
starts to move with a nonzero velocity $v$, as represented with
the red dashed line. Under this depinning regime,
\begin{equation} v\sim (E-E_{C0})^{\theta},
\end{equation}
with a velocity exponent $\theta$. Under the flow regime, when
$E\gg E_{C0}$, $v\sim E$. On the other hand, for finite $T$, the
pinning-depinning transition becomes relatively smooth, as
represented with the green solid line. Under the low $E$ (i.e.,
$\ll E_{C0}$) creep regime, the domain wall motion becomes very
slow and can be described by propagation between pinning sites due
to thermal activation. Then,
\begin{equation} v\sim exp[-(U/k_BT)(E_{C0}/E)^{\mu}],
\end{equation}
where $U$ is an energy barrier and $\mu$ is a dynamical exponent.
Critical exponents of domain dynamics, including $\mu$ and
$\theta$, can identify the universality class and provide
information on the pinning forces and the fractal nature of the
rough FE domain walls. Although measurements of these critical exponent values for the FE systems are particularly important, there are little experimental works on these values .\\
%%%%%%%%%%%%%%%%%%%%[Figure 2] %%%%%%%%%%%%%%%%%%%%%%%
\begin{figure}[tbp]
\includegraphics[width=2.4in]{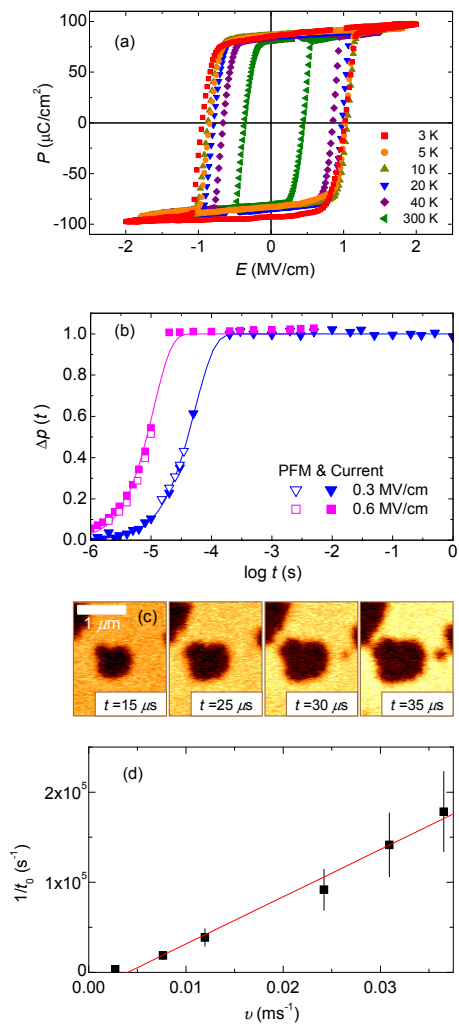}
\caption{(color online) (a) $P$-$E$ hysteresis loops at different
$T$. (b) Time-dependent normalized switched polarization, $\Delta
p(t)$, behaviors (solid symbols) from switching current and (open
symbols) from PFM measurements. The solid lines show the fitting
results using the KAI model. (c) Scanned images of $t$-dependent
domain growth at $E= 0.3$ MV/cm. (d) Plot of $1/t_0$ vs. $v$. This
relationship is linear, suggesting that $1/t_0$ from switching
current studies can be used to parameterize $v$.}
\end{figure}
%%%%%%%%%%%%%%%%%%%%%%%%%%%%%%%%%%%%%%%%%%%%%%%%%%%%%%
\indent In this Letter, we report our studies on the $T$- and
$E$-dependent nonlinear responses of FE domain wall dynamics in
epitaxial Pb(Zr,Ti)O$_{3}$ (PZT) thin film. To widen the
accessible region of $T$ and $E$, we used switching current
measurements, combined with direct $v$ data determined by PFM
images. We found that $v$ follows the nonlinear dynamic response,
as described in Fig. 1(b). We also could obtain values of the two
critical exponents, $\mu$ and $\theta$, from the data in the creep
and the depinning
regimes, respectively. This work provides us new insights on how domain walls propagate inside epitaxial FE thin films.\\
\indent We fabricated 100 nm-thick epitaxial PZT thin film on
SrRuO$_3$/SrTiO$_3$ substrate using pulsed laser deposition
\cite{Lee}. X-ray diffraction studies confirmed that a
high-quality, (001)-oriented PZT film was grown epitaxially. To
fabricate PZT capacitors, we patterned the sputtered Pt top
electrodes with a typical area of $7.5\times 10^3$ $\mu m^2$. Our epitaxial film revealed a high dielectric stability, suitable to our measurements at high electric fields.\\
\indent Figure 2(a) shows $T$-dependent polarization-electric
field ($P$-$E$) hysteresis curves for a PZT capacitor, measured
between 3 and 300 K. The saturation and remnant $P$ values were
nearly constant over a wide $T$ range. The systematic $T$
variation of the $P$-$E$ hysteresis curves comes mostly from
$T$-dependent change in coercive field $E_C$. At 3 K, $E_C$
$\approx$ 1 MV/cm. As $T$
increased, $E_C$ decreased significantly.\\
\indent One of the difficulties in performing the dynamic studies on FE domains is to measure reliable values of $v$ under uniform $E$. Recently, we developed a modified PFM technique for a FE thin film with a top metal electrode \cite{Kim, Yang}. By combining PFM with switching current measurements, we were able to track wall motions of domains. After applying one positive poling pulse (10 V, 50 $\mu$s) to pole $P$, we switched $P$ with a series of negative pulses. We then measured the PFM images after all the negative pulses. We assumed that the PFM image obtained after the negative pulses would be nearly the same as that obtained after a single pulse, with the width being equal to the sum of all the negative pulses \cite{Yang}. Typically, an image acquisition process using the PFM set-up takes a few minutes, which is relatively long in relation to the time scale corresponding to the width of the $E-$pulse used in these experiments. To check its validity, we calculated the amount of normalized switched polarization $\Delta p$ from PFM images with an area of $5 \times 5$ $\mu m^2$. These are shown as the open symbols in Fig. 2(b). We also determined $\Delta p$ independently from switching current measurements, which are shown as the solid symbols. The agreement between these data indicates that we could reliably use the modified PFM techniques to study domain wall motion in our FE film. \\
\indent Figure 2(c) shows typical PFM images of time-dependent
growth for a particular isolated domain, which were obtained at
room $T$ with $E=0.3$ MV/cm. As time $t$ increased, the domain
size increased. Eventually, it began to merge with other domains.
From the area of the isolated domain, we determined its mean
radius. By dividing the change in mean radius with the change in
$t$, we obtained the $v$ value. We repeated these procedures for
more than 20 isolated domains and finally obtained an average
value for the given $T$ and $E$. Note that our modified PFM study
can provide a way to measure $v$ directly. However, obtaining
sufficient data to plot all $v$ vs. $E$ curves would have been too
laborious. In addition, it is rather difficult to
use in the low $T$ region. Therefore, it is highly desirable to find another method to obtain $v$ reliably in a wide range of $T$ and $E$.  \\
%%%%%%%%%%%%%%%%%%%%[Figure 3] %%%%%%%%%%%%%%%%%%%%%%%
\begin{figure}[tbp]
\includegraphics[width=2.8 in]{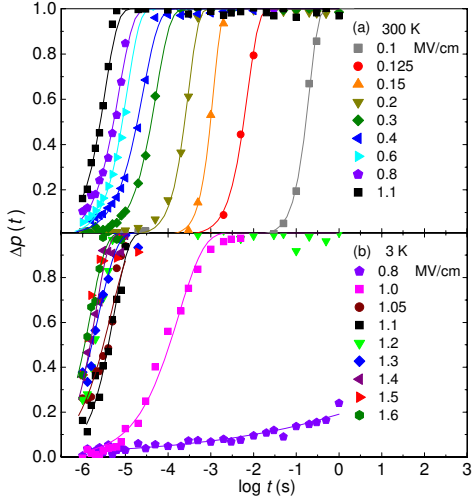}
\caption{(color online) $\Delta p(t)$ obtained via switching
current measurements under various $E$ values at (a) 300 K and (b)
3 K. The solid lines show the fitting results using the KAI
model.}
\end{figure}
%%%%%%%%%%%%%%%%%%%%%%%%%%%%%%%%%%%%%%%%%%%%%%%%%%%%%%
\indent The switching current in a FE thin film should be originated from the $P$ reversal, whose behavior is governed by domain wall dynamics.
The $t$-dependent changes in $\Delta p$ for epitaxial FE thin films \cite{So} have been explained using the Kolmogorov-Avrami-Ishibashi (KAI)
model \cite{Ishibashi}. According to the classical statistical theory on nucleation and unrestricted domain growth,
\begin{equation}
\Delta p(t) = 1 - exp[-(t/t_0)^{n}],
\end{equation}
where $n$ and $t_0$ are a geometric dimension and a characteristic
switching time for the domain growth, respectively. In the two
simplest cases, analytical relationships between $t_0$ and $v$ can
be easily obtained. When the nuclei of opposite polarity are
generated at a constant rate under $E$, $t_0\sim(1/v)^{(n-1)/n}$
\cite{Ishibashi}. Conversely, when all the nuclei are generated
instantaneously, $t_0 \sim 1/v$ \cite{Ishibashi}. Our previous
studies on epitaxial PZT films showed that nucleation rate is
approximately proportional to $1/t$ \cite{Kim}, which is much
closer to the latter case, so $1/t_0$ might be nearly proportional
to $v$.\\
\indent We investigated the relationship between $v$ and $1/t_0$
using the switching current response. Our experimental $\Delta
p(t)$ data were fitted using the KAI model, i.e. Eq. (3), as
displayed by the solid lines in Fig. 2(b). This prediction was
closely correlated with the $\Delta p(t)$ data. We compared
$1/t_0$ values with $v$ values, which were directly measured by
PFM. As shown in Fig. 2(d), $1/t_0$ is
linearly proportional to $v$ with a small offset, indicating that the switching current response could provide reliable values of $v$.\\
%%%%%%%%%%%%%%%%%%%%[Figure 4] %%%%%%%%%%%%%%%%%%%%%%%
\begin{figure}[tbp]
\includegraphics[width=3.0in]{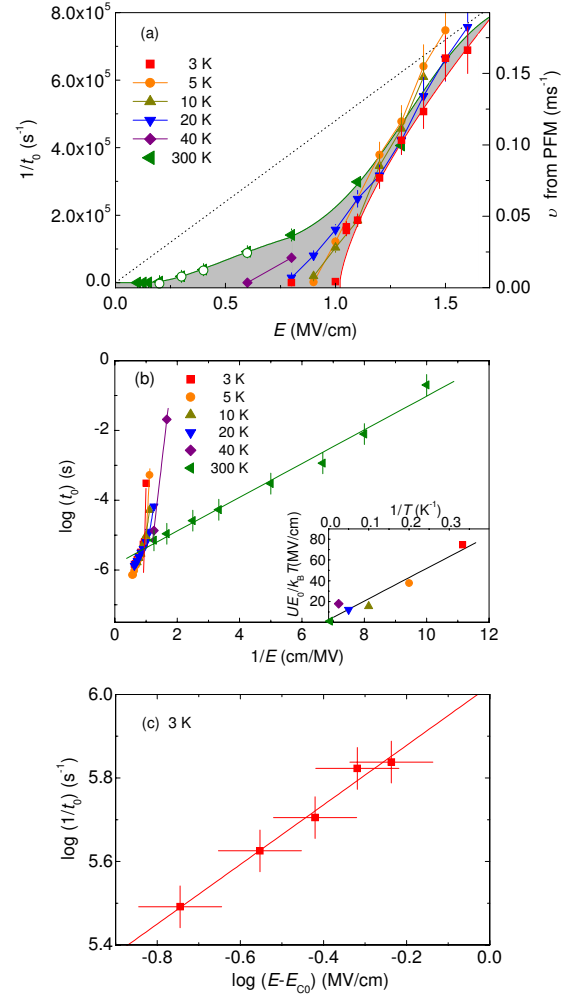}
\caption{(color online) (a) (Solid symbols) $T$-dependent curves
of $1/t_0$ vs. $E$. The dotted and solid lines are guidelines for
eye. (b) $log(t_0)$ vs. $1/E$ curves. The linear fitting indicates
a dynamic exponent $\mu \sim 1$. The inset shows that
$UE_{C0}/k_B$ is nearly independent of $T$. (c) Plot of $log
(1/t_0)$ vs. $log (E-E_{C0})$ with the $E_{C0} \sim 1$ MV/cm.
From the slope of the plot, we found a velocity exponent $\theta
\sim 0.7$.}
\end{figure}
%%%%%%%%%%%%%%%%%%%%%%%%%%%%%%%%%%%%%%%%%%%%%%%%%%%%%%
\indent Figure 3 shows $\Delta p(t)$ data with varying values of
$E$. The solid symbols and lines indicate the experimental data
and the fitting results using Eq. (3), respectively. At 300 K, a
sudden change in $\Delta p(t)$ with $E=0.1$ MV/cm occurred between
0.1 and 1.0 s: namely, at $t_0 \sim 2 \times 10^{-1}$ s. As $E$
increased, a sudden change in  $\Delta p(t)$ occurred over a
shorter timescale, i.e., a smaller value of $t_0$. Conversely, at
3 K, little changes were observed in $\Delta p(t)$ when $E$ was
below 0.8 MV/cm. This implied that the FE domain was pinned by
defects and could not move below a threshold value of $E$.
However, $\Delta p(t)$ starts to change abruptly around $E=1$
MV/cm. This corresponded to the pinning-depinning transition. When
$E$ increased, $t_0$ rapidly decreased. The $\Delta p(t)$ data for
3 and 300 K were similar at $E=1.1$ MV/cm; $t_0 \sim 5.4 \times
10^{-6}$ and $3.3 \times 10^{-6}$ s at 3 and
300 K, respectively. These similar $t_0$ values demonstrated that the $T$-dependence of $v$ became insignificant under very high $E$ region.\\
\indent Figure 4(a) shows experimental $1/t_0$ vs. $E$ curves. Note that these $1/t_0 - E$ plots resemble those in
Fig. 1(b). In the low $E$ region, $1/t_0$ was strongly dependent on
$T$, consistent with predictions for the thermally activated creep
regime. However, in the high $E$ region, the values of $1/t_0$
started to merge, indicating a crossover to the flow regime. The
pinning-depinning transition at 3 K occurred at approximately 1 MV/cm, which was close to the $E_C$ value obtained from $P-E$ hysteresis curve in Fig. 2(a).\\
\indent From the experimental $1/t_0$ data, we obtained the value
of the dynamical exponent $\mu$ for the creep regime. From Eq.(2),
$log(t_0)$ should be proportional to $(U/k_BT)(E_{C0}/E)^{\mu}$.
The $log(t_0)$ vs. $1/E$ curves are shown in Fig. 4(b). The
experimental data at a given $T$ falls approximately into a linear
line, indicating that $\mu$ is close to 1.0. Considering
experimental errors, we found $\mu = 0.9 \pm 0.1$. This $\mu$
value agrees with previously reported values for epitaxial and
polycrystalline PZT films \cite{Tybell,Jo}. The inset shows the
value of $UE_{C0}/k_BT$ calculated from lines of best fit for
several values of $T$. The straight line indicates that the value
of $UE_{C0}/k_B$ is nearly independent of $T$ and is about 300
K$\cdot$MV/cm, close to previously reported value of around 400
K$\cdot$MV/cm from local domain switching data using PFM at room
$T$ \cite{Tybell}.\\
\indent The value of $\mu$ reflects the nature of the pinning
potential in our PZT films. Under a pinning potential with a short
range (the so-called random bond), one- and two- dimensional
domain walls should have the $\mu$ values of 0.25 and 0.5,
respectively \cite{Chauve}. However, under a pinning potential
with a long range (the random field), $\mu= 1.0$ regardless of
dimensionality \cite{Chauve}. Recently, two conflicting $\mu$
values were reported for epitaxial PZT thin films; 1.0 and 0.5-0.6
\cite{Tybell,Paruch}. Our study confirms that $\mu = 0.9 \pm 0.1$
over wide range of $T$. This $\mu$ value suggests that the defects
in our PZT thin film induce a long-ranged local field and
pin FE domain walls \cite{remark}.\\
\indent From our experimental $1/t_0$ data, we also obtained the
value of the velocity exponent $\theta$ near the pinning-depinning
transition. By Eq.(1), $log(1/t_0)$ should be proportional to
$\theta \cdot log(E-E_{C0})$. As shown in Fig. 4(c), this is
approximately the case when $T=3$ K. The slope of the line for
best fit was $\theta \approx 0.71 \pm 0.05$. We could not find any
earlier studies with which compare this $\theta$ value for FE thin
films.\\
\indent The $\theta$-value should reflect the dimensionality $D$
of the surface for an elastic object in a disordered medium, as
$\theta=(5+D)/9$ \cite{Chauve}. Using this relationship, we found
that $D \approx 1.4\pm 0.4$ for our PZT thin film. Recently,
several studies have reported non-integer dimensionality of local
and static domain walls in FE thin films \cite{Rodriguez,Catalan}.
By measuring the local switching of PZT-BiFeO$_3$ sol-gel thin
films with liquid PFM, Rodriguez $et$ $al.$ reported that
$D\approx 1.5\pm0.1$ \cite{Rodriguez}. By measuring morphology and
scaling of the domains in epitaxial BiFeO$_3$ thin films, Catalan
$et$ $al.$ reported that $D\approx 1.5\pm 0.1$ \cite{Catalan}. Our
measured values of $\theta$ from the dynamic responses is
consistent with these $D$ values from the static measurements. In
addition, our findings indicate that the newly observed fractal
dimensionality
of FE domains should be originated from local quenched defects.\\
\indent We want to point out future studies and possible
implications of our works: (i) in order to fully understand the
domain wall dynamics in epitaxial FE thin films, we need to
measure other critical exponents, related to the divergence of the
correlation length, the local variance of the domain wall
position, and so on \cite{Barabasi}; (ii) domain dynamic responses
in systems with different structural conditions such as
polycrystalline FE thin films and ultrathin films should be
investigated and compared with those in epitaxial films; (iii) the
complete understanding on FE domain dynamics is necessary for
numerous practical applications, including optimizing operation
speed of miniaturized FE devices.\\
 \indent In summary, we
investigated domain wall motions of epitaxial PZT film over a wide
range of temperature and applied electric field. We found that the
motions were likely to be governed by the nonlinear dynamics of
surface growth in a disordered medium with quenched defects. We
determined two critical exponents for domain wall propagation
dynamics, which indicate the random field nature of the defects
and fractal nature of domain walls. Our works provide us a new
future direction of studies on the domain wall motions in
ferroelectric materials.\\
\indent We acknowledge valuable discussions with B. Khang. This
study was financially supported by the Creative Research
Initiatives (Functionally Integrated Oxide Heterostructures) of
the Ministry of Science and Technology (MOST), the Korean Science
and Engineering Foundation (KOSEF), and the laboratory Directed
Research and Development Program of Oak Ridge National Laboratory
(H.N.L). J.Y.J. acknowledges
the financial support, in part, of Brain Korea 21.\\


\begin{references}
\bibitem{Sornette} D. Sornette, \textit{Critical Phenomena in Natural Sciences}
(Springer, Berlin, 2003).
\bibitem{Barabasi} A.-L. Barab\'{a}si and H.E. Stanley, \textit{Fractal Concepts in Surface Growth} (Cambridge University Press, Cambridge, 1995).
\bibitem{Dawber} M. Dawber, K.M. Rabe, and J.F. Scott, Rev. Mod.
Phys. \textbf{77}, 1083 (2005).
\bibitem{Tagantsev} A.K. Tagantsev and G.
Gerra, J. Appl. Phys. \textbf{100}, 051607 (2006).
\bibitem{Kim} D.J. Kim \textit{et al.}, Appl. Phys. Lett. \textbf{91}, 132903 (2007).
\bibitem{Rodriguez} B.J. Rodriguez \textit{et al.}, Phys. Rev. Lett. \textbf{98}, 247603
(2007).
\bibitem{Lee} H.N. Lee \textit{et al.}, Phys. Rev. Lett. \textbf{98}, 217602
(2007).
\bibitem{Yang} S.M. Yang \textit{et al.}, Appl. Phys. Lett. \textbf{92}, 252901 (2008).
\bibitem{So} Y.W. So \textit{et al.}, Appl. Phys. Lett. \textbf{86}, 092905
(2005).
\bibitem{Ishibashi} Y. Ishibashi and Y. Takagi, J. Phys. Soc.
Jpn. \textbf{31}, 506 (1971).
\bibitem{Tybell} T. Tybell \textit{et al.},
Phys. Rev. Lett. \textbf{89}, 097601 (2002).
\bibitem{Jo} J.Y. Jo \textit{et al.}, Phys. Rev.
Lett. \textbf{99}, 267602 (2007).
\bibitem{Chauve} P. Chauve, T. Giamarchi, and P. Le Doussal, Phys. Rev. B \textbf{62}, 6241 (2000).
\bibitem{Paruch}    P. Paruch, T.
Giamarchi, and J.-M. Triscone, Phys. Rev. Lett. \textbf{94},
197601 (2005).
\bibitem{remark} Electrostatic calculations resulting in a value of $\mu =1$ have also been obtained for a periodic lattice potential without defects. However, based on electrostatic calculations for domain wall motion in a periodic potential, it has been reported that the calculated values for the activation field (i.e., $UE_0/k_BT$) and critical length of nuclei disagree with experimental values for thin films
\cite{Tybell}.
\bibitem{Catalan} G. Catalan \textit{et al.}, Phys. Rev.
Lett. \textbf{100}, 027602 (2008).
\end{references}
\end{document}